\documentstyle[12pt]{article}

\begin{document}
\baselineskip=20.5pt

\def\beqra{\begin{eqnarray}} \def\eeqra{\end{eqnarray}}
\def\beqast{\begin{eqnarray*}} \def\eeqast{\end{eqnarray*}}
\def\beq{\begin{equation}}	\def\eeq{\end{equation}}
\def\be{\begin{enumerate}}   \def\ee{\end{enumerate}}

\def\fnote#1#2{\begingroup\def\thefootnote{#1}\footnote{#2}\addtocounter
{footnote}{-1}\endgroup}

\def\ut#1#2{\hfill{UTTG-{#1}-{#2}}}
\def\fl#1#2{\hfill{FERMILAB-PUB-94/{#1}-{#2}}}

\def\sppt{Research supported in part by the
Robert A. Welch Foundation and NSF Grant PHY 9009850}

\def\utgp{\it Theory Group\\ Department of Physics \\ University of Texas
\\ Austin, Texas 78712}

\def\gam{\gamma}
\def\Gam{\Gamma}
\def\la{\lambda}
\def\eps{\epsilon}
\def\La{\Lambda}
\def\si{\sigma}
\def\Si{\Sigma}
\def\al{\alpha}
\def\Tha{\Theta}
\def\tha{\theta}
\def\vphi{\varphi}
\def\del{\delta}
\def\Del{\Delta}
\def\ab{\alpha\beta}
\def\om{\omega}
\def\Om{\Omega}
\def\mn{\mu\nu}
\def\mun{^{\mu}{}_{\nu}}
\def\kap{\kappa}
\def\rsi{\rho\sigma}
\def\beal{\beta\alpha}

\def\til{\tilde}
\def\rta{\rightarrow}
\def\eqv{\equiv}
\def\nab{\nabla}
\def\pa{\partial}
\def\sit{\tilde\sigma}
\def\ul{\underline}
\def\indt{\parindent2.5em}
\def\nd{\noindent}

\def\rsi{\rho\sigma}
\def\beal{\beta\alpha}

\def\caa{{\cal A}}
\def\cb{{\cal B}}
\def\cac{{\cal C}}
\def\cd{{\cal D}}
\def\ce{{\cal E}}
\def\cf{{\cal F}}
\def\cg{{\cal G}}
\def\cah{{\cal H}}
\def\ci{{\cal I}}
\def\cj{{\cal{J}}}
\def\ck{{\cal K}}
\def\cl{{\cal L}}
\def\cm{{\cal M}}
\def\cn{{\cal N}}
\def\cO{{\cal O}}
\def\cp{{\cal P}}
\def\car{{\cal R}}
\def\cs{{\cal S}}
\def\ct{{\cal{T}}}
\def\cu{{\cal{U}}}
\def\cv{{\cal{V}}}
\def\cw{{\cal{W}}}
\def\cx{{\cal{X}}}
\def\cy{{\cal{Y}}}
\def\cz{{\cal{Z}}}

\def\raisenot{\raise .5mm\hbox{/}}
\def\nota{\ \hbox{{$a$}\kern-.49em\hbox{/}}}
\def\notA{\hbox{{$A$}\kern-.54em\hbox{\raisenot}}}
\def\notb{\ \hbox{{$b$}\kern-.47em\hbox{/}}}
\def\notB{\ \hbox{{$B$}\kern-.60em\hbox{\raisenot}}}
\def\notc{\ \hbox{{$c$}\kern-.45em\hbox{/}}}
\def\notd{\ \hbox{{$d$}\kern-.53em\hbox{/}}}
\def\notbd{\ \hbox{{$D$}\kern-.61em\hbox{\raisenot}}} 
\def\note{\ \hbox{{$e$}\kern-.47em\hbox{/}}}
\def\notk{\ \hbox{{$k$}\kern-.51em\hbox{/}}}
\def\notp{\ \hbox{{$p$}\kern-.43em\hbox{/}}}
\def\notq{\ \hbox{{$q$}\kern-.47em\hbox{/}}}
\def\notW{\ \hbox{{$W$}\kern-.75em\hbox{\raisenot}}}
\def\notz{\ \hbox{{$Z$}\kern-.61em\hbox{\raisenot}}}
\def\notpa{\hbox{{$\partial$}\kern-.54em\hbox{\raisenot}}}

\def\fo{\hbox{{1}\kern-.25em\hbox{l}}}  
\def\rf#1{$^{#1}$}
\def\bx{\Box}
\def\tr{{\rm Tr}}
\def\rmtr{{\rm tr}}
\def\dgg{\dagger}

\def\lag{\langle}
\def\rag{\rangle}
\def\bmid{\big|}

\def\vlap{\overrightarrow{\La p}} 
\def\lrta{\longrightarrow} \def\lrar{\raisebox{.8ex}{$\longrightarrow$}}
\def\rlarw{\longleftarrow\!\!\!\!\!\!\!\!\!\!\!\lrar}

\def\llra{\relbar\joinrel\longrightarrow}              
\def\mapright#1{\smash{\mathop{\llra}\limits_{#1}}}    
\def\mapup#1{\smash{\mathop{\llra}\limits^{#1}}}     

\def\7#1#2{\mathop{\null#2}\limits^{#1}}	
\def\5#1#2{\mathop{\null#2}\limits_{#1}}	
\def\too#1{\stackrel{#1}{\to}}
\def\tooo#1{\stackrel{#1}{\longleftarrow}}
\def\nout{{\rm in \atop out}}

\def\one{\raisebox{.5ex}{1}}
\def\BM#1{\mbox{\boldmath{$#1$}}}

\def\ltsim{\matrix{<\cr\noalign{\vskip-7pt}\sim\cr}}
\def\gtsim{\matrix{>\cr\noalign{\vskip-7pt}\sim\cr}}
\def\haf{\frac{1}{2}}


\def\place#1#2#3{\vbox to0pt{\kern-\parskip\kern-7pt
                             \kern-#2truein\hbox{\kern#1truein #3}
                             \vss}\nointerlineskip}

\def\illustration #1 by #2 (#3){\vbox to #2{\hrule width #1 height 0pt depth
0pt
                                       \vfill\special{illustration #3}}}

\def\scaledillustration #1 by #2 (#3 scaled #4){{\dimen0=#1 \dimen1=#2
           \divide\dimen0 by 1000 \multiply\dimen0 by #4
            \divide\dimen1 by 1000 \multiply\dimen1 by #4
            \illustration \dimen0 by \dimen1 (#3 scaled #4)}}

\def\ON{{\cal O}(N)}
\def\UN{{\cal U}(N)}
\def\bdPh{\mbox{\boldmath{$\dot{\!\Phi}$}}}
\def\bPh{\mbox{\boldmath{$\Phi$}}}
\def\bPhs{\bPh^2}
\def\sef{S_{eff}[\sigma]}
\def\sigx{\sigma(x)}
\def\bph{\mbox{\boldmath{$\phi$}}}
\def\bphs{\bph^2}
\def\ex{\BM{x}}
\def\exs{\ex^2}
\def\xdot{\dot{\!\ex}}
\def\y{\BM{y}}
\def\ys{\y^2}
\def\ydot{\dot{\!\y}}
\def\pat{\pa_t}
\def\pax{\pa_x}

\renewcommand{\thesection}{\Roman{section}}
\ut{05}{94}

\fl{255}{T}

\hfill{hep-th/9408120}

\vspace*{.3in}
\begin{center}
  \large{\bf On Kinks and Bound States in the Gross-Neveu Model}
\normalsize

\vspace{36pt}
Joshua Feinberg\fnote{*}{ Supported by a post doctoral Rothchild Fellowship
and in part by the Robert A. Welch Foundation and NSF Grant PHY 9009850.}

\vspace{12pt}
{\it Theory Group, Department of Physics\\
The University of Texas at Austin, RLM5.208, Austin, Texas 78712\\
and \\
Theory Group, Fermi National Accelerator Laboratory\\
P.O. Box 500, Batavia, IL 60510\\
\vspace{4pt}
e-mail joshua@utaphy.ph.utexas.edu}

\vspace{.6cm}

\end{center}

\begin{minipage}{5.3in}
{\abstract~~~~~
We investigate static space dependent $\sigx=\lag\bar\psi\psi\rag$ saddle
point configurations in the two dimensional Gross-Neveu model in the large N
limit. We solve the saddle point condition for $\sigx$ explicitly by employing
supersymmetric quantum mechanics and using simple properties of the diagonal
resolvent of one dimensional Schr\"odinger operators rather than inverse
scattering techniques. The resulting solutions in the sector of unbroken
supersymmetry are the Callan-Coleman-Gross-Zee kink configurations. We thus
provide a direct and clean construction of these kinks. In the sector of broken
supersymmetry we derive the DHN saddle point configurations. Our method of
finding such non-trivial static configurations may be applied also in other two
dimensional field theories.}

\end{minipage}

\vspace{48pt}
PACS numbers: 11.10.Lm, 11.15.Pg, 11.10.St, 11.10.Kk

\vfill
\pagebreak

\setcounter{page}{1}

\section{Introduction}

The Gross-Neveu model \cite{gn} is a well known two dimensional field
theory of $N$ massless Dirac fermions $\psi_a\,(a=1,\ldots,N)$ with
$\UN$ invariant self interactions, whose action is given by
\beq
S=\int d^2x\left[\sum_{a=1}^N\, \bar\psi_a\,i\notpa\,\psi_a +
\frac{g^2}{2}\; \left( \sum_{a=1}^N\;\bar\psi_a\,\psi_a\right)^2
\right]\;.
\label{1}
\eeq
We are interested in the large N limit of (\ref{1}) in which $N\rightarrow
\infty$ while $Ng^2$ is held fixed. Decomposing each Dirac spinor into two
Majorana spinors one observes that $S$ is invariant under $\cO(2N)$ flavour
symmetry containing the $\UN$ mentioned above as a subgroup \cite{dhn}. The
field theory defined by (\ref{1}) is a renormalisable field theory exhibiting
asymptotic freedom, dynamical symmetry breaking and dimensional transmutation.
Its spectrum was calculated semiclassically (in the large $N$ limit) in
\cite{dhn}. It contains the fermions in (\ref{1}) which become massive, as well
as a rich collection of bound states thereof (the so called "DHN states"). The
spectrum of (\ref{1}) contains also kink configurations \cite{ccgz,klein}.
We refer to these as the Callan-Coleman-Gross-Zee (CCGZ) kinks in the sequel.
These kinks are expected to be part of the spectrum of the Gross-Neveu model
since dynamical breaking of the discrete chiral symmetry in the Gross-Neveu
model suggests that there should be extremal field configurations that
interpolate between the two minima of the {\sl effective\/} potential
associated with (\ref{1}) in much the same way that such configurations arise
in classical field theories whose potential term has two or more equivalent
minima.

The Gross-Neveu model has also a system of infinitely many (non-local)
conservation laws which (presumably) forbid particle production in scattering
processes and enables the exact calculation of $S$-matrix elements in the
various sectors of the model \cite{soliton}. Results of such calculations
are in agreement with the ``Large N" calculation of the spectrum.

In this paper we discuss static space dependent $\sigx=\lag\bar\psi\psi
\rag$ configurations that are solutions of the saddle point equation governing
the effective action corresponding to (\ref{1}) as $N\rightarrow\infty$. Such
$\sigx$ configurations correspond to non-trivial excitations of the vacuum
\cite{cjt,jg} and are therefore important in determining the entire spectrum
of the field theory in question\cite{dhn} and its finite temperature behaviour
as well\cite{finitet}. Such configurations are important also in discussing the
behaviour of the ${1\over N}$ expansion of (\ref{1}) at large orders
\cite{bh,dev,br}.

Our discussion makes use of supersymmetric quantum mechanics and simple
properties of the diagonal resolvent of one dimensional Schr\"odinger
operators. Using these two basic tools we are able to solve the saddle point
condition for static $\sigx$ configurations explicitly.
The supersymmetry alluded to above relates the upper and lower components of
spinors, implying that the square of the Dirac operator may be decomposed into
two isospectral Schr\"odinger operators. It is closely related to the soliton
degeneracy discussed in \cite{jr}, where the soliton is considered as a
degenerate doublet having fermion numbers $\pm\haf$. Our explicit solution
of the saddle point condition in the sector of unbroken supersymmetry consists
of the CCGZ kink configurations and it therefore provides a clean and direct
construction of these kinks. In the sector of broken supersymmetry, our
explicit
solution reproduces the so called DHN saddle point configurations\cite{dhn}.

In a recent paper \cite{jf}, we have applied a similar method to
the anharmonic $\ON$ oscillator and the two dimensional $\ON$ vector model in
the limit $N\rightarrow\infty$. In the latter case we have found that the
effective action sustains in the large temperature limit
extremal bilinear condensates of the $\ON$ vector field that are analogous to
the CCGZ kinks in the Gross-Neveu model.

The paper is organised as follows: In Section II we analyse the saddle point
equation for static $\sigx$ configurations employing supersymmetric quantum
mechanics. Using simple manipulations, we show that the latter equation may be
expressed in terms of the hamiltonian of only one of the supersymmetric
sectors.
In Section III.1 we resolve the saddle point equation into frequencies and
demand that the extremum condition be satisfied by each Fourier component
separately.
This strong condition turns out to leave us always in the sector of unbroken
supersymmetry. This corresponds physically to $\sigx$ configurations that
interpolate between the two vacua of the Gross-Neveu model at the two ends of
the world, which are the CCGZ kinks that we find as explicit solutions of the
saddle point equation. In Section III.2 we solve the saddle point equation
without
separating it into frequencies, under the assumption that there is only one
bound state. This leads to the DHN $\sigx$ configurations which belong to the
sector of broken supersymmetry. We conclude our discussion in section IV.

\pagebreak

\section{The Saddle Point Equation for Static Solutions}

Following\cite{gn} we rewrite (\ref{1}) as
\beq
S=\int d^2x\,\left[i\bar\psi\notpa\psi-g\si\bar\psi \psi-\haf\,\si^2
\right]
\label{2}
\eeq
where $\si(x)$ is an auxiliary field \cite{fn1}.

Thus, the partition function associated with (\ref{2}) is
\beq
\cz=\int\,\cd\si\,\cd\bar\psi\,\cd\psi \,\exp\left\{
i\int\limits^{\infty}_{-\infty}dt\int\limits^{\infty}_{-\infty}dx\left[
\bar\psi
i\notpa\psi-g\si\bar\psi\psi-\haf\,\si^2\right]\right\}
\label{4}
\eeq
Gaussian integration over the grassmannian variables is
straightforward, leading to $\cz=\int\,\cd\si\,\exp \{iS_{eff}[\si]\}$
where the bare effective action is \cite{fn2}
 \beq
S_{eff}[\si] =-\haf\int\limits^{\infty}_{-\infty} dt \int\limits^{\infty}_{-
\infty} dx \si^2-i\frac{N}{2}\, \tr\ln\left[-(i\notpa-g\si)(i\notpa +g\si)
\right]\,.
\label{5}
\eeq
The ground state of the Gross-Neveu model (\ref{1}) is described by the
simplest extremum of $S_{eff}$\cite{gn} in which $\si=\si_0$ is a constant that
is fixed by the (bare) gap equation
\beq
-g\si_0 + iNg^2\,{\rm tr}\int
{d^2k\over\left(2\pi\right)^2}{1\over\notk-g\si_0}
= 0\,.
\label{bgap}
\eeq
Therefore, the dynamically generated mass of small fluctuations of the Dirac
fields around this vacuum is
\beq
m = g\si_0 = \mu\,e^{1-{\pi\over Ng_R^2\left(\mu\right)}}
\label{mass}
\eeq
where $\mu$ is an arbitrary renormalisation scale, and the renormalised
coupling $g_R\left(\mu\right)$ is related to the cut-off dependent bare
coupling via $ \Lambda\,e^{-{\pi\over Ng^2\left(\Lambda\right)}} = \mu\,e^{1-
{\pi\over Ng_R^2\left(\mu\right)}}$ where $ \Lambda$ is an ultraviolet cutoff.
Since $m$ is the physical mass of the fermions it must be a renormalisation
group invariant, and this fixes the scale dependence of the renormalised
coupling $g_R$ ,namely, equation (\ref{mass}). From now on we will drop the
subscript $R$ from the renormalised coupling and simply denote it by $Ng^2$.

As was explained in the introduction, we are interested in more complicated
extrema of $S_{\it eff}$, namely static space dependent solutions of the
extremum condition on $S_{\it eff}$. This condition reads generally
\begin{eqnarray}
&&\frac{\del S_{eff}}{\del\si(x,t)} = -\si(x,t)
\nonumber \\
&& -i\frac{N}{2} \rmtr
\left\{\left[ 2g^2\si(x,t)+ig\gam^\mu\pa_\mu\right]\lag x,t|\, [\,\bx +g^2
\si^2-ig\gam^\mu\pa_\mu\si]^{-1}|x,t\rag\right\}=0
\label{7}
\end{eqnarray}
where ``tr'' is a trace over Dirac indices.

Specialising to static $\sigx$ configurations, and using the Majorana
representation  $\gam^1=i\si_3\;,\; \gam^0=\si_2$ for $\gam$ matrices,
(\ref{7}) becomes
\beqra
&&{2i\over Ng}\si(x)=\,\rmtr \left[\left(\begin{array}{cc} 2g\si-\pa_x &
0 \\ 0 & 2g\si+\pa_x \end{array}\right)\int\limits^{\infty}_{-\infty}\,
{d\om\over 2\pi}\left(
\begin{array}{ll}R_+\left(x,\om^2\right) &~~~~0 \\ ~~~~0 &R_-\left(x,\om^2
\right)\end{array}\right)\right]\,=
\nonumber \\
&&{}
\nonumber \\
&&\int\limits^{\infty}_{-\infty}\,{d\om\over 2\pi}\,\left[\left(2g\si-
\pa_x\right)R_+\left(x,\om^2\right)+\left(2g\si+\pa_x\right)R_-\left(x,\om^2
\right)\right]
\label{8}
\eeqra
where
\beq
R_{\pm}(x,\om^2)=
\lag x|\frac{1}{h_{\pm}-\om^2}\,| x\rag
\label{9}
\eeq
 are the diagonal resolvents of the one dimensional Schr\"odinger operators
\beq
h_{\pm}=-\pa_x^2+g^2\si^2\pm g\si'(x)
\label{10}
\eeq
evaluated at spectral parameter $\om^2$.\\

Note that $h_{\pm}$ are positive semidefinite isopectral (up to zero-modes)
hamiltonians, since (\ref{10}) may be brought into the form
\cite{witt,rosner}
\begin{eqnarray}
&&h_+=Q^{\dagger}Q ~,~ h_-=QQ^{\dagger}~~~~where
\nonumber \\
&& Q=-{d\over dx}+g\si ~ , ~ Q^{\dagger}={d\over dx}+g\si .
\label{12}
\end{eqnarray}
These operators may be composed into a supersymmetric hamiltonian
$H=\left(\begin{array}{cc} h_+ & 0 \\ 0 & h_- \end{array}\right)$
describing one bosonic and one fermionic degrees of freedom \cite{witt}, which
we identify with the upper and lower components of the spinors.
Supersymmetry implies here that an interchange of the bosonic and fermionic
sectors of $H$ leaves dynamics unchanged, as can be seen from the fact that
Eqs. (\ref{8})-(\ref{12}) are invariant under the simultaneous interchanges
\begin{eqnarray}
&&\si\rightarrow -\si~~,~~ h_{\pm}\rightarrow h_{\mp},
\nonumber \\
&& R_{\pm}\rightarrow R_{\mp}.
\label{13}
\end{eqnarray}

Isospectrality of $h_+$ and $h_-$ alluded to above means that to each
eigenvector $\psi_n$ of $h_-$ with a {\it positive} eigenvalue $E_n$,
there is a corresponding eigenvector $\phi_n$ of $h_+$ with the same
eigenvalue and norm, and vice-versa. The precise form of this pairing relation
is
\beqra
&&\phi_n\,=\, {1\over\sqrt{E_n}}Q^{\dagger}\psi_n,
\nonumber\\
&&\psi_n\,=\,{1\over\sqrt{E_n}}Q\phi_n\quad ;\quad  E_n>0\, .
\label{iso}
\eeqra
It is clear that the pairing in (\ref{iso}) fails when $E_n=0$. Thus, in
general one cannot relate the eigenvectors with zero-eigenvalue (i.e.-the
normalisable zero-modes) of one hamiltonian in (\ref{10}) to these of the
other.
Should such a normalisable zero-mode appear in the spectrum of one of the
positive semidefinite operators in (\ref{10}), it must be the ground state
of that hamiltonian. In this case the lowest eigenvalue of the supersymmetric
hamiltonian H is zero, which is the case of unbroken supersymmetry. If such a
normalisable zero-mode does not appear in the spectrum, all eigenvalues of H,
and in particular-its ground state energy, are positive, and supersymmetry is
broken. Since the ground state of a Schr\"odinger operator is non-degenerated,
$h_{\pm}$ can have each no more than one such a normalisable zero-mode.
Moreover, it is clear from (\ref{10}) that in our one dimensional case, only
{\it one} operator in (\ref{10}) may have a normalisable zero-mode, since it
must be annihilated either by $Q$ or by $Q^{\dagger}$. In cases of unbroken
supersymmetry, we will take such a normalisable zero-mode to be an eigenstate
of $h_-$, namely, the {\it real} function
\beq
\Psi_0\left(x\right)\,=\,\cn\,e^{-g\int\limits^{x}_{0}\si\left(y\right)dy}
\label{zeromode}
\eeq
which is the normalisable solution of the differential equation
\beq
Q^{\dagger}\Psi_0\,=\,0
\label{qdagger}
\eeq
where $\cn$ is a normalisation coefficient.

Note that a necessary condition for the normalisability of $\Psi_0$ is that
$\sigx$ have the opposite behaviour at $\pm\infty$. Thus, physically, cases of
unbroken supersymmetry lead to $\sigx$ configurations that interpolate between
the two vacua of (\ref{5}) at the two ends of the world, while cases of broken
supersymmetry yield $\sigx$ configurations that leave and return to the same
vacuum state.

Assigning the zero-mode to $h_-$ poses no loss of generality, since the other
possible case is related to this one via (\ref{13}). In what follows we will
denote the right hand side of (\ref{zeromode}) by $\Psi_0$ also in cases of
broken supersymmetry where it is non-normalisable. This should not cause any
confusion, since the ground state will be denoted by $\psi_0$, which will be
equal to $\Psi_0$ when supersymmetry is unbroken.

By definition, the diagonal resolvents in (\ref{9}) are given by the
eigenfunction expansions
\beqra
&&R_-\left(x\right)\,=\,\sum^{\infty}_{n=0}\,{|\psi_n\left(x\right)|^2\over
{E_n-\om^2}}\, ,
\nonumber\\
&&R_+\left(x\right)\,=\,\sum^{\infty}_{n>0}\,{|\phi_n\left(x\right)|^2
\over{E_n-\om^2}}
\label{resolvents}
\eeqra
where the sums extend over all eigenstates, including the continua of
scattering states where they are understood as integrals.

Using (\ref{iso}), $R_+$ may be expressed in terms of the $\psi_n$'s as
\beq
R_+(x,\om^2)\,=\,\sum^{\infty}_{n>0}\, {1\over E_n\Psi_0^2\left(x
\right)}   {|W_n\left(x\right)|^2\over {E_n-\om^2}}
\label{r+}
\eeq
where $W_n(x)\,=\,\Psi_0(x)\psi_n^{'}(x)\,-\,\Psi_0^{'}(x)\psi_n(x)$ is
the wronskian of $\Psi_0$ and $\psi_n$. An elementary consequence of the
Schr\"odinger equation is that $W_n^{'}(x)\,=\,-E_n\Psi_0(x)\psi_n(x)$. Using
this relation, (\ref{resolvents}) and (\ref{r+}) imply the important relation
\beq
\left(2g\si-\pa_x\right)R_+\,+\,\left(2g\si+\pa_x\right)R_-\,=\,2\left(2g\si
+{d\over dx}\right)\lag x|\,\cp{1\over h_- - \om^2}\,| x\rag\, .
\label{important}
\eeq
Here $\cp$ is the projector
\beq
\cp\,=\,{\bf 1}-\lambda|\psi_0\rangle\langle\psi_0|
\label{projector}
\eeq
that projects out the ground state of $h_-$ when supersymmetry is unbroken
($\lambda = 1$), and is just the unit operator otherwise ($\lambda = 0$).
We can also use (\ref{bgap}) to make a frequency resolution of unity
as
\beq
{i\over Ng^2}=\int\limits^{\Lambda}_{-\Lambda}\, {d\om\over\pi}\,\lag x|
{1\over -\pa_x^2+m^2-\om^2-i\epsilon}|x\rag\,.
\label{resolution}
\eeq
Moreover, from (\ref{important}), (\ref{resolution})  and the elementary
relation
\beq
\lag x|{1\over -\pa_x^2+m^2-\om^2-i\epsilon}|x\rag\,=\,{1\over
2\sqrt{m^2-\om^2-i\epsilon}}\,,
\label{free}
\eeq
the frequency resolution of (\ref{8}) becomes
\beqast
{d\over dx}\lag x|\cp {1\over h_- - \om^2} |x\rag\, =\, 2g\sigx\left[
{1\over2\sqrt{m^2-\om^2}} - \lag x|\cp {1\over h_- -\om^2} |x\rag\right]\, .
\eeqast

Substituting (\ref{projector}) into the last equation, all dependence on
$\lambda$ cancels out and the extremum condition (\ref{8}) is shaped into its
final form
\beq
\int\limits^{\infty}_{-\infty}\,{d\om\over 2\pi}\,\left\{
{d R_-\left(x,\om^2\right)\over dx}\, -\, 2g\sigx\left[
{1\over 2\sqrt{m^2-\om^2}}\,-\,R_-\left(x,\om^2\right)\right]\right\}
\label{static'}
\eeq
which has to be satisfied regardless of whether supersymmetry is broken or not.

\pagebreak

\section{Static Solutions to the Extremum Condition}

\subsection{Case of Unbroken Supersymmetry}

The simplest way to look for solutions of the static saddle point equation
(\ref{static'}) is to demand that it be satisfied by each frequency mode
separately, namely restricting $R_-$ by the differential condition

\beq
{d R_-\left(x,\om^2\right)\over dx}\, =\, 2g\sigx\left[
{1\over 2\sqrt{m^2-\om^2}}\,-\,R_-\left(x,\om^2\right)\right] \,.
\label{static}
\eeq

Now, $R_-\left(x,\om^2\right)$ , being the diagonal resolvent of $h_-$ at
spectral parameter $\om^2$ is subjected to the so-called "Gel'fand-Dikii"
equation\cite{gd}\cite{fn6}
\beq
-2R_-(x,\om^2)R_-^{\prime\prime}(x,\om^2)+\left(R_-^{\prime}(x,\om^2)\right)^2+
4R_-^2(x,\om^2)\left[g^2\si^2\,-\,g\si^{\prime}-\om^2\right]=1  .
\label{11}
\eeq

Therefore, both equations (\ref{static}) and (\ref{11}) must hold and they form
a system of coupled non-linear differential equations in the unknowns
$\sigx,~{\rm and}~R_-(x,\om^2)$. Substituting $R_-^{\prime}$ and
$R_-^{\prime\prime}$ from (\ref{static}) into (\ref{11}) we obtain a quadratic
equation for $R_-$ whose solutions are

\beq
R_-\left(x,\om^2\right)\,=\,{-g\si^{\prime}\,\pm\,\sqrt{\left(
g\si^{\prime}\right)^2 + 4\om^2g^2\si^2 - 4\om^2\left(m^2-\om^2\right)}\over
4\om^2\sqrt{m^2-\om^2}}\,.
\label{r-}
\eeq

To see what the two signs of the square root correspond to we observe that
the solution with the negative sign in front of the square root has a simple
pole as a function of $\om^2$ at $\om^2=0$ with a negative residue, while the
other solution is regular and positive at $\om^2=0$. Therefore, from
(\ref{resolvents}) it is clear that the negative sign root corresponds to the
case of unbroken supersymmetry, where the simple pole signals the existence
of a normalisable zero-mode in the spectrum of $h_-$, while the positive root
solution corresponds, for similar reasons, to cases in which $h_-$ lacks such a
zero-mode. We will see below that the latter solution corresponds
also to the case of unbroken supersymmetry, where the zero-mode is in the
spectrum of $h_+$.

We concentrate now on the branch cut singularity in (\ref{r-}) at
$\om^2=m^2$ which signals the threshold of the continuous part of the
spectrum of $h_-$. Clearly, $h_-$ can have no continuous spectrum
other than that starting at $\om^2=m^2$, thus leading to the conclusion that
$R_-$ in (\ref{r-}) can have no other branch points in the $\om^2$ plane.
Therefore, the expression under the square root in (\ref{r-}) must be a perfect
square as a polynomial in $\om^2$, namely,
\beq
\pm\,g\si^{\prime}(x)=g^2\si^2(x)-m^2\,.
\label{square}
\eeq
{}From (\ref{square}) we find straight forwardly the solutions
\beq
g\sigx=\pm\,m\,tanh\left[m\left(x-x_0\right)\right]
\label{ccgz}
\eeq
which are exactly the CCGZ kinks and anti-kinks. Here the parameter $x_0$ is an
integration constant that implies translational invariance of (\ref{ccgz})
since it is the arbitrary location of the kinks. Clearly, both cases in
(\ref{ccgz}), and therefore both cases in (\ref{r-}), lead to $h_{\pm}$
operators that do not break supersymmetry, since (\ref{zeromode}),(\ref{ccgz})
imply that
\beq
\Psi_0 = \left(m\over 2\right)^{\frac{1}{2}}e^{-g\int_0^x\si\left(y\right)dy} =
\left(m \over 2\right)^{\frac{1}{2}}sech[m(x-x_0)]
\label{ofan}
\eeq
is the normalisable zero-mode of $h_-$ for the kink configuration, and of $h_+$
when $\sigx$ is the anti-kink. Due to zero binding energy, fermions trapped in
this potential do not react back on the $\sigx$ field \cite{dhn}.

Note that in deriving (\ref{ccgz}) above we have
not set any apriori restrictions on $R_-$, thus, it is a very interesting
question whether (\ref{ccgz}) are the {\sl only\/} possible static extremal
$\sigx$ configurations in the sector of unbroken supersymmetry or not.

As it stands, our
frequency decomposition of the extremum condition (\ref{static}) seems to lead
always to extremal $\sigx$ configurations that do not break supersymmetry.
We are thus unable to find in this manner the extremal $\sigx$ configurations
found in \cite{dhn} in which $\sigx$ has the shape of kink anti-kink pair that
are very close to each other. Such a configuration evidently breaks
supersymmetry \cite{witt}. They will be discussed in the next subsection.

When $\sigx$ is a kink (\ref{r-}) becomes
\beq
R_-(x,\om^2)= -{m^2~sech^2\left[m\left(x-x_0\right)\right]\over
2\om^2\sqrt{m^2-\om^2}}+{1\over 2\sqrt{m^2-\om^2}}
\label{rkink}
\eeq
while for anti-kinks it is just the expression on the right hand side of
(\ref{free}).

These statements on $R_-$ are consistent with the explicit form of the
hamiltonians $h_{\pm}$. Using (\ref{10}),(\ref{ccgz}) we find in the kink case
\beqra
&&h_+ = -\pa_x^2 + g^2\si^2 + g\si^{\prime} = -\pa_x^2 + m^2 \quad {\rm and}
\nonumber\\
&&h_- = -\pa_x^2 + g^2\si^2 - g\si^{\prime} = -\pa_x^2 + m^2 - 2m^2\,sech^2
[m(x-x_0)]
\label{reflectionless}
\eeqra
while in the anti-kink case $h_{\pm}$ interchange their roles .
Thus, in the latter case, $h_-$ becomes the Schr\"odinger operator of a freely
moving particle in a {\sl constant\/} potential $m^2$ which is the reason why
$R_-$ is given by the simple expression in (\ref{free}) in the anti-kink case.
Indeed, that expression is the $x$ independent solution of the Gel'fand-Dikii
equation (\ref{11}) corresponding to the constant potential $g^2\si^2-g\si^
{\prime}=m^2$ that is positive at $\om^2 = 0$.

In the kink case, we have obtained the explicit form (\ref{rkink}) for $R_-$
by substituting (\ref{ccgz}) into (\ref{r-}). As an independent check, we may
deduce this expression for the diagonal resolvent for the potential $g^2\si^2-
g\si^{\prime}\,=\,m^2-2m^2\, sech[m(x-x_0)]$ in other ways. The simplest one is
to apply an ansatz of the form $R_- = \alpha\, sech^2(\beta x) + \gam$ to the
Gel'fand-Dikii equation. Another way to derive it is to use the general formula
$R(x) = G(x,y)|_{x=y} = {\psi_1(x_>)\psi_2(x_<) \over W(\psi_1,\psi_2)}|_{x=y}$
where the $\psi$'s are the so called Jost functions of the problem. In this
case they are hypergeometric functions multiplied by $sech$ factors. The
wronskian in this expression is a ratio of $\Gamma$ functions dependent on
$\om^2$ and $m^2$\cite{jg}.

We can deduce from (\ref{r-}) the CCGZ solution (\ref{ccgz}) to the extremum
condition (\ref{static}),(\ref{11}) by yet another method which does not rely
upon the brach-cut structure of $R_-$ but rather on its pole structure.
Considering the case where the zero-mode is in the spectrum of $h_-$ we make
a Laurent expansion of the appropriate expression for $R_-$ in (\ref{r-})
around the simple pole at $\om^2=0$. The leading term in this expansion is
\beq
R_-\,\sim\, -{1\over \om^2}\left[{g\si^{\prime}\over 2m}\,+\,\cO\left(\om^2
\right)\right]\,.
\label{pole}
\eeq
Comparing (\ref{pole}) to (\ref{zeromode}) and (\ref{resolvents}) we find
\beq
\Psi_0^2(x)\,=\,\cn^2 exp\left(-2g\int\limits^{x}_{0}\si(y) dy\right)\,=\,
{g\si^{\prime}\over 2m}
\label{kimat}
\eeq
which yields
\beq
\cn^2e^{-2\phi}={\phi^{\prime\prime}\over 2m}\quad ;\quad \phi(x)=g\int
\limits^{x}_{0}
\si(y) dy \,.
\label{diffeq}
\eeq

Solving (\ref{diffeq}) we find
\beq
\phi(x)\,=\,ln\,cosh[m(x-x_0)] + c
\label{phi}
\eeq
where $c$ and $x_0$ are integration constants and we have imposed the
normalisation condition $\int\limits^{\infty}_{-\infty}\,\Psi_0^2\,dx=1$ to fix
$\cn=\sqrt{{m \over 2c}}$. Clearly, then, (\ref{phi}) yields the CCGZ kinks
upon differentiation.

We are now in a position to verify briefly that the CCGZ kink configuration
leads to an $h_-$ operator that has indeed a single normalisable zero-mode as
the explicit form (\ref{rkink}) for $R_-$ suggests. Our discussion follows
\cite{rosner}. In the kink sector (\ref{reflectionless}) implies that the
eigenstates of $h_+$ are simply these of freely moving particles $\phi_{k}(x) =
e^{ikx}$, with a {\sl continuum \/} of strictly positive eigenvalues $E_+=k^2+
m^2\geq m^2$. These states are isospectral to the eigenstates of $h_-$,
\beq
\psi_{k}(x)={1\over\sqrt{k^2+m^2}}Q\,\phi_{k}(x)=\left\{\,{m\,tanh\,[m(x-x_0)]\,
-ik\over \sqrt{k^2+m^2}}\,\right\}\,e^{ikx}
\label{15}
\eeq
which are therefore the {\sl scattering \/} states of $h_-$. The $S$ matrix
associated with $h_-$ is thus
\beq
S(k)={ik-m\over ik+m}=exp\left[i\left(\pi-2~arctan{k\over m}\right)\right] .
\label{16}
\eeq
Since $h_+$ in (\ref{reflectionless}) has only scattering states, $h_-$ can
have no bound states other than its zero-mode (\ref{ofan}) which must therefore
be its ground state. This {\sl single \/} normalisable state of $h_-$
corresponds to the single pole of $S(k)$ in (\ref{16}) at $k=im$.
Note further that there are no reflected waves in any of the scattering
eigenstates (\ref{15}) of the Schr\"odinger operators in (\ref{reflectionless})
. This is also the case for the supersymmetry breaking $\sigx$ configurations
in \cite{dhn} as well as in other exactly soluble models in two space-time
dimensions\cite{dhn}.

The fact that $h_{\pm}$ evaluated at the extremal point must be reflectionless
can be deduced even without solving the extremum condition (\ref{static})
explicitly, provided one makes apriori an assumption that $h_-$ has only a
single bound state (namely, its ground state) regardless of whether
supersymmetry is broken or not.

To this end we consider (\ref{static}), in which $R_-$ may be replaced by
$R_P\,=\,\lag x|\,\cp{1\over h_- - \om^2}\,| x\rag\,$ as mentioned in the
discussion preceding (\ref{static}). Since $h_-$ is assumed to have a single
bound state, $R_P$ contains only scattering states of $h_-$, whose
corresponding continuous eigenvalues $E=k^2+m^2$ start at $E=m^2$ . We may
therefore write the spectral resolution of $R_P$ as
\beq
R_P = \int\limits^{\infty}_{m^2} dE {|\psi_E\left(x\right)|^2\over E-\om^2} =
\int\limits^{\infty}_{-\infty} dk {\rho_k\left(x\right)\over k^2+m^2-\om^2}
\label{rp}
\eeq
where $\rho_k\left(x\right) = 2\pi|k||\psi_k\left(x\right)|^2$. In terms of
$\rho_k$ the extremum condition (\ref{static}) becomes
\beq
{d\over dx}{\rho_k\left(x\right)\over \Psi_0^2\left(x\right)} = {d\over dx}{1
\over \Psi_0^2\left(x\right)}
\label{staticrho}
\eeq
whose general solution is
\beq
\rho_k\left(x\right) = 1 + c_k\Psi_0^2\left(x\right)
\label{solrho}
\eeq
where $c_k$ is an integration constant.
Since by definition $\rho_k\left(x\right)$ cannot blow up at infinity, we must
set $c_k=0$ in the case of broken supersymmetry. Whether supersymmetry is
broken or not $\rho_k\left(x\right)$ obviously obtains the asymptotic value of
$1$ as $x\rightarrow\pm\infty$. Therefore, the scattering states
$\psi_k\left(x\right)$ of $h_-$ are given by
\beq
\psi_k\left(x\right) = \sqrt{1 + c_k\Psi_0^2\left(x\right)}\,\,
e^{i\alpha_k\left(x
\right)}
\label{scatteringstates}
\eeq
where $\alpha_k$ is a real phase. Substituting these functions into the
eigenvalue equation for $h_-$ and considering its asymptotic behaviour as
$x\rightarrow\pm\infty$ we see, using the boundary condition $g\si\left(\pm
\infty\right) = \pm m$ that the phase becomes that of a free particle, which is
obvious, but unimodularity of the phase factor implies further that there be
only right moving or only left moving wave in $\psi_k\left(x\right)$.
Therefore, $h_-$ must be reflectionless.

The physical significance of the CCGZ kinks is as follows:
As was mentioned in the introduction, the dynamical properties
of (\ref{1}) are consequences of the fact that the ("large N") effective
potential $V_{{\rm eff}}(\si)$ extracted from (\ref{5}) has two symmetric
equivalent minima at $\lag\si\rag_{{\rm vac}}=\pm\si_0\neq 0$. This causes a
dynamical breakdown of the discrete ($Z_2$) chiral symmetry of (\ref{2}) under
the transformation $\psi\rightarrow \gam_5\psi,\,\si\rightarrow-\si\,$, where
the fermions fluctuating near the $\lag\si\rag_{{\rm vac}}=\pm\si_0$ ground
state acquire dynamical mass $m=\pm g\si_0$\cite{gn}. In a similar manner to
the appearance of kinks in classical field theories with potentials exhibiting
spontaneous symmetry breaking, one should expect similar configurations to
appear in field theories whose {\it effective} potential implies dynamical
symmetry breaking. The CCGZ kinks (anti-kinks) are precisely such static space-
dependent $\sigx$ configurations that interpolate between the two minima of
$V_{{\rm eff}}(\si)$, and our calculations provide an explicit proof that they
are indeed extremal points of (\ref{5}). The various states appearing in the
background of the CCGZ kink may be deduced by calculating the "partition
function" of the Dirac field fluctuations in that specific $\sigx$
configuration for a finite time lapse $T$ (i.e.-the trace over the time
evolution operator $e^{-iHT}$) . This has been done explicitly in
\cite{dhn}. The result is
\beqra
&&\tr\,e^{-iHT} = \sum_{n=0}^{2N} {\left(2N\right)!\over \left(n\right)!
\left(2N-n\right)!}\, exp\left[-{i\over 2}\int\limits^{T}_{0} dt
\int\limits^{\infty}_{-\infty} dx \left(\si_{kink}^2-\si_0^2\right)\right]\cdot
\nonumber\\
&&exp\left\{iNT\left(\sum_i\om_i\left[\si_{kink}\right] - \sum_i\om_i\left[
\si_0\right]\right) -in\om_b\left[\si_{kink}\right]T\right\}\,.
\label{partition}
\eeqra
In this equation $n$ is the total number of fermions and anti-fermions that are
trapped in the single bound state of the kink, $\om_i\left[\si\right] =
\sqrt{E_i}$ is the energy of the $i-th$ state in the background of
$\si$\cite{fnenergy} and $\om_b$ is the energy of the bound state, which
is zero for the CCGZ kinks. Using this and also the fact that in this
background $h_{\pm}$ in (\ref{reflectionless}) are isospectral, we see that all
terms in the second exponent in (\ref{partition}) are cancelled, leaving
only the first exponent which is the mass of the kink,
namely\cite{fnpi},
\beq
M_{kink} = \haf \int\limits^{\infty}_{-\infty}\left[\si_0^2 - \si_{kink}^2
\right] = {mN\over Ng^2}\,.
\label{kinkmass}
\eeq

Therefore, we see that all states contributing to (\ref{partition}) are
degenerate in energy (all having the kink mass as energy) which is a direct
result of the fact that $\om_b=0$. Clearly there are $2^{2N} $ states in all,
that form a huge reducible supermultiplet of $\cO(2N)$. Its decomposition into
irreducible components is clear from the combinatorial prefactors in
(\ref{partition}) that simply tell us that the various bound states in the kink
fall into anti-symmetric tensor representations of $\cO(2N)$  (where the
integer $n$ is the rank of the tensor)\cite{dhn}.

We close this subsection by checking explicitly that the CCGZ kink
configurations obtained above indeed extremise the effective action in
(\ref{5}).
Substituting the kink configuration in (\ref{ccgz}) and the explicit
expressions (\ref{rkink}) and (\ref{free}) for $R_-$ and $R_+$ into the
extremum condition (\ref{8}) we find that the pole at $\om^2=0$ disappears from
the right hand side of (\ref{8}) in accordance with (\ref{important}) leaving
in the sum over frequencies only contributions from the scattering states.
Thus, Eq.(\ref{8}) becomes
\beq
\left[1+iNg^2\int\limits^{\Lambda}_{-\Lambda}{d\om\over 2\pi}
{1\over\sqrt{m^2-\om^2 - i\epsilon}}\right]g\sigx=0  ,
\label{20}
\eeq
implying that the term in the square brackets on the left hand side must
vanish. But vanishing of the latter is precisely the Minkowsky space
gap-equation of the Gross-Neveu model (\ref{bgap}) for the dynamical mass
$m$ and it must therefore hold, confirming that the kinks in (\ref{ccgz}) are
indeed solutions of the extremum condition (\ref{8}).

\pagebreak

\subsection{Case of Broken Supersymmetry}

Enforcing the saddle point condition at each frequency component in
(\ref{static}) led us directly to the sector of unbroken supersymmetry without
any further assumptions on $R_-(x,\om^2)$. Therefore, in order to find static
extremal $\sigx$ configurations that lead to supersymmetry breaking, we must
solve (\ref{static'}) as a whole. At a first sight this seems to be
unmanageable\cite{fndhn}, since we apparently cannot use the Gel'fand-Dikii
equation (\ref{11}) for $R_-$ which was so crucial for our treatment in the
previous section. However, assuming (as in \cite{dhn}) that the required
$\sigx$ configuration yields an $h_-$ operator with a {\sl single} bound state
at {\sl positive} energy $E_b=\om_b^2<m^2$, in addition to the obvious
continuum of unbound fermions of mass $m$, our experience gained in the
previous section leads us to the most general form of
$R_-(x,\om^2)$ consistent with (\ref{11}) and the preassumed form of the
spectrum of $h_-$. This generic form of $R_-$ contains enough information in
order to solve (\ref{static'}).

In order to construct this generic form of $R_-$, we note the following points:
\begin{enumerate}
\item $R_-$ has scale dimension $-1$ in mass units as can be seen from its
definition (\ref{9}) or from the explicit expressions (\ref{r-}) and
(\ref{rkink}).
\item Clearly, $g\sigx$ must reach asymptotically either one of the vacua,
namely, $|g\si(\pm\infty)| = m\,,\,g\si'(\pm\infty)=0$ (actually, since
we will end up indeed in the sector of broken supersymmetry, the two asymptotic
values of $g\si$ will turn out to be equal). Thus, the $h_-$ operator
resulting from such a $\sigx$ configuration must have the asymptotic form
$-\pa_x^2+m^2$ as $|x|\rightarrow\infty$. Correspondingly, its resolvent must
have the asymptotic form (\ref{free}).
\item The assumption that $h_-$ has a single bound state at positive energy
$0<\om_b^2<m^2$ implies that the analytic structure of $R_-(x,\om^2)$ in the
complex $\om$ plane must include in addition to the branch cuts along the real
rays $(-\infty,-m)$ and $(m,\infty)$ only two simple poles on the real axis at
$\om=\pm\om_b$ .
\item Any ansatz for $R_-$ must obey the Gel'fand-Dikii equation (\ref{11}).
\end{enumerate}

Points number $1$ and number $3$ lead immediately to the general form
\beq
R_-(x,\om^2) = {A g\si' + B \left(g\si\right)^2 + C\left(\om_b^2-\om^2\right)
+ D m^2 \over \left(\om_b^2-\om^2\right)\sqrt{m^2-\om^2}}
\label{ansatz}
\eeq
where $A,B,C$ and $D$ are dimensionless numbers yet to be determined.
Point number $2$ imply then that $C={1\over 2}$ and $B=-D$.

In order to comply with point number $4$, we substitute (\ref{ansatz}) into
the Gel'fand-Dikii equation (\ref{11}). Multiplying the resultant expression
through by $(\om_b^2-\om^2)^2 (m^2-\om^2)$ we obtain an even quartic polynomial
in $\om$\cite{fnquartic} which must vanish. This yields three non-linear
differential equations for $g\sigx$ stemming from nullification of the
coefficients of $\om^4, \om^2$ and $\om^0$ in that polynomial. The condition
which results from setting the coefficient of $\om^4$ to zero reads
\beq
(4A-1) (g\si)' + (1-4D) [(g\si)^2 - m^2]=0
\label{quartic}
\eeq
which can be used to eliminate $A$ and $D$ from (\ref{ansatz}). Doing so
(\ref{ansatz}) becomes
\beq
R_-(x,\om^2)= { g\si' + m^2 - \left(g\si\right)^2
\over 4\left(\om_b^2-\om^2\right)\sqrt{m^2-\om^2}} + {1\over 2\sqrt{m^2-\om^2}}
\,.
\label{generalform}
\eeq
This expression is evidently very similar to expressions (\ref{r-}) and
(\ref{rkink}), the only difference being that the double pole at $\om=0$
in those equations is resolved here into the two simple poles at
$\om=\pm\om_b$.

We must further subject (\ref{generalform}) to the two remaining differential
equations mentioned above. It turns out that the equation resulting from
setting the coefficient of $\om^2$ to zero is simply the derivative (with
respect to $x$) of the equation associated with the coefficient of $\om^0$
\cite{fngelfand}. The latter is the differential equation for $g\sigx$ we are
looking for, which must be solved subjected to the boundary conditions
mentioned in point number $2$ above. Equivalently, and this is what we do
below,
we can
read off the expression for the {\sl normalised} bound state wave function
$\psi_b(x)$ in terms of $g\sigx$ from (\ref{generalform}). The Schr\"odinger
equation for $\psi_b(x)$ provides than the required condition for $g\sigx$.

As can be seen from (\ref{resolvents}) and in a similar manner to our
discussion following (\ref{pole}) in the previous subsection, we identify the
residue of the simple pole of $R(x,E\equiv\om^2)$ in (\ref{generalform}) at
$E=\om_b^2$ as $-\psi_b^2(x)$. Thus,
\beq
\psi_b(x)= \sqrt{ {g\si'-\left(g\si\right)^2+m^2 \over 4 \sqrt{m^2-\om_b^2}}}
=\sqrt{{m^2-V \over 4 \sqrt{m^2-\om_b^2}}}
\label{psib}
\eeq
where
\beq
V(x)=g^2\si^2-g\si'
\label{potential}
\eeq
is the potential of $h_-$. Imposing the Schr\"odinger equation on (\ref{psib})
we have
\beq
\left[-\pa_x^2 + V(x)\right]\sqrt{m^2-V(x)}=\om_b^2 \sqrt{m^2-V(x)}\,.
\label{vequation}
\eeq
The solution to this equation, compatible with boundary conditions at
infinity is
\beq
V(x)=m^2-2\kappa^2 sech^2 [\kappa (x - x_0)]
\label{v}
\eeq
where $\kappa^2 \equiv m^2-\om_b^2$ and $x_0$ is an integration constant. The
corresponding bound state and resolvent are therefore
\beqra
&&\psi_b(x)=\sqrt{{\kappa \over 2}} sech [ \kappa (x-x_0 )]
\nonumber\\
&&R_-(x,\om^2)= { \kappa^2 sech^2 [\kappa (x - x_0)
\over 2\left(\om_b^2-\om^2\right)\sqrt{m^2-\om^2}} + {1\over 2\sqrt{m^2-\om^2}}
\label{dhnr-}
\eeqra
which solve (\ref{11}) and (\ref{vequation}) as can be checked explicitly. It
is not surprising at all that (\ref{dhnr-}) reproduces (\ref{rkink}) as
$\om_b\rightarrow 0$.

Finally, in order to find $g\sigx$ we substitute (\ref{v}) into
(\ref{potential}). This is equivalent to solving
\beq
\left\{-\pa_x^2+m^2-2\kappa^2 sech^2 [\kappa (x - x_0)]\right\} \Psi_0 (x)=0
\label{predhn}
\eeq
where $\Psi_0(x) $ is defined in terms of $g\sigx$ in (\ref{zeromode}).
For $\kappa^2\neq m^2$ the Schr\"odinger operator in (\ref{predhn}) has no
normalisable zero-mode and supersymmetry is broken. Nevertheless, one can
find non-normalisable solutions of the differential equation (\ref{predhn}),
which can be transformed into an hypergeometric equation \cite{jg},
and extract $g\sigx$ in this way. The resulting $g\sigx$ configurations are
those found in \cite{dhn}. The specific $g\sigx$ given in \cite{dhn}
corresponds to setting $\kappa x_0 = {1\over 4 } ln {m + \kappa \over
m - \kappa }$ in (\ref{v}), namely,
\beq
g\sigx = m + \kappa\left\{ tanh\left[ \kappa x - {1\over 4 } ln
{m + \kappa  \over m - \kappa }\right] - tanh\left[ \kappa x + {1\over 4 }
ln {m + \kappa  \over m - \kappa }\right]\right\}
\label{dhnconfig}
\eeq

Note at this stage that we have yet to impose the saddle point condition
(\ref{static'}). This will quantise $\om_b$ and constrain it to the discrete
set of values found by DHN \cite{dhn} as we now show.

Substituting (\ref{generalform}) into the generic static saddle point condition
(\ref{static'}) we obtain
\beq
\left[\left(2g\si + {d\over dx}\right)\left(g\si'+m^2-g^2\si^2\right)
\right]\int_{{\cal C},n} {d\om\over 2\pi}
{1 \over \left(\om_b^2-\om^2\right)\sqrt{m^2-\om^2}}\,=\,0
\label{finally}
\eeq
where we now integrate over $\om$ along a contour $\cac$ in the complex
$\om$ plane, and the subindex $n$ counts the number of fermions trapped in the
single bound state of $h_-$ produced by $g\si$. The contour $\cac$ is
precisely the one used
to define the Feynman propagator of a free Dirac particle of mass $m$, and in
addition, it runs right {\sl below both} simple poles of $R_-$ at
$\om=\pm\om_b$\cite{fncontour}.

The resulting static saddle point condition (\ref{finally}) is solved by
requiring either that the term in the square brackets vanishes, or that the
integral over frequencies vanishes. The differential equation resulting from
the
first possibility is
\beq
g\si''=2g\si (g^2\si^2 - m^2)
\nonumber\\
\eeq
which reproduces the kink configuration (\ref{ccgz}) discussed in the previous
section with $\om_b=0$. We thus focus on the other possibility, namely, that
the integral in (\ref{finally}) vanishes.

The integral in (\ref{finally}) is U.V.
finite. We can therefore deform it by folding its $\om>\om_b$ wing right on top
of the remaining part of the real $\om$ axis to the left of $\om_b$. This can
be further deformed into two circles wrapped around the simple poles at
$\pm\om_b$ and a ``hair-pin" configuration wrapped around the left hand cut
of $R_-$, picking up its discontinuity across the cut\cite{fncontour}. In other
words, this integral picks up contributions
from the completely filled Dirac sea (including the pole at $\om=-\om_b$),
where
each energy state (of the Dirac operator $i\notpa-g\si$ rather than $h_-$)
is occupied by N fermion flavours, and the single bound state at $\om=+\om_b$,
which is occupied by $n<N$ fermions.

Because we have pulled out a factor of
$N$ in deriving (\ref{static'}) we have to weigh the contribution of the simple
pole of $R_-$ at $\om=\om_b$ by $n/N$ in (\ref{finally}) while the other
contributions are weighed simply by $1$ . Performing the
integration, we find that the contribution from the cut is
$-{i\over m^2 sin 2\theta} (1-{2\theta \over \pi})$ while the poles at
$\om=\pm\om_b$ yield, respectively,  $\mp{i\over m^2 sin 2\theta}$ where,
following \cite{dhn} we have defined
\beqra
&&\om_b=m~ sin \theta~~~~~ {\rm and}
\nonumber\\
&&\kappa=\sqrt{m^2-\om_b^2}=m~ cos\theta .
\label{theta}
\eeqra

Gathering all contributions to the integral (properly weighed), (\ref{finally})
yields
\beqast
\int_{{\cal C},n} {d\om\over 2\pi}
{1 \over \left(\om_b^2-\om^2\right)\sqrt{m^2-\om^2}}\,=\,
{i\over m^2 sin 2\theta}\left[ {2\theta\over \pi} - {n\over N}\right]\,=\,0
\eeqast
namely,
\beq
\theta_n = {\pi n \over 2N}
\label{dhn}
\eeq
which is precisely the result of \cite{dhn}.
The simple poles in $R_-$ occur therefore at
\beq
\om_b=m~ sin \left({\pi n \over 2N}\right)\,.
\label{omb}
\eeq

One can now repeat the same analysis as in the previous subsection, of the
partition function associated with fluctuations of fermions interacting with
these static $\sigx$ configurations.  One finds again $\cO (2N)$
supermultiplets, where the n'th supermultiplet has mass $M_n = {2Nm\over \pi}
sin\left({\pi n\over 2N}\right),~ (n<N) $ and contains all $\cO (2N)$
completely antisymmetric tensors of rank $n_0\leq n $, where $n_0$ has the same
parity of $n$\cite{dhn}.
\pagebreak

\section{Conclusion}

In this paper we have solved the extremum condition on the effective action
of the two dimensional Gross-Neveu model for {\it static} $\sigx$
configurations in the large N limit. Our method of calculation was direct,
making use of elementary properties of one dimensional
Schr\"odinger operators. The natural scale that appears in the Gross-Neveu
model is that of its dynamically generated mass $m$. Because the latter is
the natural scale of the CCGZ kink configurations as well, our derivation of
these kink configurations was relatively simple and straight forward. It
therefore may be considered as a clean and constructive proof that the CCGZ
kinks are indeed static extrema of the effective action. We have also
rederived the DHN extremal $\sigx$ configurations in a very simple manner.
To this end, however, we had to introduced their scale into the saddle point
condition by hand, since it does not appear explicitly in the effective action.

Our method may be applied also to a host of
other two dimensional field theories, and in particular, to field theories
that {\sl do not\/} involve reflectionless static configurations, where
inverse scattering methods are useless\cite{deVega}.\\
{}\\

I would like to thank the Theory Group at Fermilab where part of this work has
been completed for their kind hospitality. I would also like to thank Joe
Lykken for inviting me to Fermilab and for valuable discussions on topics
related to this work. I am also indebted to the Theory Group at the University
of Texas at Austin for partial support of this visit.

\pagebreak

{\bf Note Added in Proof:}
After submitting this paper for publication I have realised that
Avan and de Vega \cite{deVega} have already used the diagonal resolvent of a
one dimensional Schr\"odinger operator to discuss solitons in ``Large N" vector
models. However, their discussion was limited to single particle quantum
mechanics ($0+1$ dimensional quantum field theory). Moreover, they did
not make an explicit use of the Gel'fand-Dikii equation as was done here.
I thank the referee for bringing this reference to my attention and for his
useful remarks.
\pagebreak


\begin{thebibliography}{99}
\bibitem{fn1} From this point to the end of this paper flavour indices
are suppressed.  Thus $i\bar\psi\notpa\psi$ should be understood as
$\displaystyle{i\sum^N_{a=1} \bar\psi_a \notpa\psi_a}$.  Similarly
$\bar\psi\psi$ stands for $\displaystyle{\sum^N_{a=1} \,\bar\psi_a\psi_a}$.
\bibitem{fn2}  Here we have used the fact that $D_1=i\notpa-g\si$ and
$D_2=-(i\notpa+g\si)$ are isospectral (up to zero-modes) since
$\gam_5D_1=D_2\gam_5$\,.
\bibitem{fnenergy} Note that we used the term "energy" in refering to
eigenvalues of the Schr\"odinger operators $h_{\pm}$, but clearly $\om_i$ is
really the physical energy of the fermion.
\bibitem{fnpi} It is customary to choose the renormalisation point at $\mu=m$
such that $Ng^2=\pi$.
\bibitem{fndhn} Indeed, the original analysis in \cite{dhn} was carried out
using sophisticated inverse scattering techniques rather than a direct
calculation of the resolvent of the static Dirac operator. See also the
comments made by these authors in \cite{dhn1} (p. 4136, first paragraph
starting below fig.2) prior to their analysis in \cite{dhn}.
\bibitem{fnquartic} Note that we multiply through by a polynomial of degree six
in $\om$. However, the coefficient of $\om^6$ in the resulting polynomial
vanishes identically.
\bibitem{fngelfand} This result is a special case of the recurrence relations
discussed in \cite{gd}.
\bibitem{fncontour}
That is , $\cac$ runs right below the left branch cut of $R_-$ in the
complex $\om$ plane, continuing right below the real $\om$ axis up
to the right most pole at $\om=\om_b$ and then crossing the real $\om$ axis
and continuing to run right above the right branch cut of $R_-$.
This is precisely the contour used in \cite{dhn1}. See figures $1$ through $3$
(and especially the last two figures) in that reference.
\end{thebibliography}
\end{document}